\documentclass[twocolumn,aps,epsfig,nofootinbib]{revtex4}

\usepackage{graphicx}
\usepackage{epstopdf}
\usepackage{latexsym}
\usepackage{amssymb}
\usepackage{amsmath}
\usepackage{color}
\usepackage{mathrsfs}
\usepackage[center]{subfigure}

\begin{document}

\newcommand{\bq}{\begin{equation}}
\newcommand{\eq}{\end{equation}}
\newcommand{\bqn}{\begin{eqnarray}}
\newcommand{\eqn}{\end{eqnarray}}
\newcommand{\nb}{\nonumber}
\newcommand{\lb}{\label}
\newcommand{\PRL}{Phys. Rev. Lett.}
\newcommand{\PL}{Phys. Lett.}
\newcommand{\PR}{Phys. Rev.}
\newcommand{\PRD}{Phys. Rev. D}
\newcommand{\CQG}{Class. Quantum Grav.}
\newcommand{\JCAP}{J. Cosmol. Astropart. Phys.}
\newcommand{\JHEP}{J. High. Energy. Phys.}
\newcommand{\PLB}{Phys. Lett. B}

\title{Quantization of 2d  Ho\v{r}ava gravity: non-projectable case}

\author{Bao-Fei Li$^{a, b}$}
\email{Bao-Fei$\_$Li@baylor.edu}

\author{V. H. Satheeshkumar${}^{c}$}
\email{vhsatheeshkumar@gmail.com}
 
\author{Anzhong Wang$^{a, b, c}$\footnote{Corresponding author}}
\email{Anzhong$\_$Wang@baylor.edu}

\affiliation{$^{a}$ Institute  for Advanced Physics $\&$ Mathematics, Zhejiang University of Technology, Hangzhou 310032,  China\\
$^{b}$ GCAP-CASPER, Physics Department, Baylor University, Waco, TX 76798-7316, USA\\
$^{c}$  Departamento de F\'{\i}sica Te\'orica, Instituto de F\'{\i}sica, UERJ, 20550-900, Rio de Janeiro, Brazil}

\date{\today}

\begin{abstract}

The quantization of two-dimensional  Ho\v{r}ava theory of gravity without the projectability condition is considered. Our study of the  
Hamiltonian structure of the theory shows that there are two first-class and two second-class constraints.  Then, following Dirac we 
quantize the theory by first  requiring that the two second-class constraints be strongly equal to zero. This is carried out by replacing 
the  Poisson  bracket by the Dirac bracket. The two first-class  constraints give rise to the Wheeler-DeWitt equations, which yield 
uniquely a plane-wave solution for the wavefunction.  We also study the classical solutions of the theory and find that the characteristics  
of classical spacetimes  are encoded solely in the phase of the  plane-wave solution in terms of the extrinsic curvature of the foliations 
$t =$Constant, where $t$ denotes the globally-defined time of the theory. 

\end{abstract}

 
\maketitle
 
\section{Introduction}

The study of two dimensional quantum gravity has been a favorite playground for theoretical physicists who are engaged in reconciling the principles of quantum mechanics with gravity at high energies. Given the insurmountable difficulties one faces when attempting to quantize gravity in the usual four dimensions, it is pragmatic to try to gain some insights by studying the lower dimensional models which are simpler yet share some interesting features with the four dimensional theory. This kind of approach has helped us earlier in understanding the other three fundamental interactions of nature. In fact, the dynamical gauge symmetry breaking was first understood in the two dimensional model of quantum electrodynamics due to Schwinger \cite{Schwinger:1962tp} and the large-N behavior of non-Abelian gauge theories was first found in the solution of two dimensional quantum chromodynamics \cite{'tHooft:1974hx}. Even the modern approach to string theory was born with Polyakov's observation \cite{Polyakov:1981rd} that a first-quantized string propagating in \textit{d}-dimensional flat target space can be described as a theory of \textit{d} free scalar fields coupled to two dimensional quantum gravity \cite{Polyakov:1987zb}. So one hopes that quantizing gravity in two dimensions is a worthwhile exercise in order to understand its four dimensional counterpart. See the references \cite{Jackiw:1984je, Brown:1988,GKV} for  review of such efforts.


One of the reasons that 2d quantum gravity became so popular  is that it is a theory on the world-sheet of both ``critical" as well as ``non-critical" string theories. Following Polyakov \cite{Polyakov:1981rd}, two different approaches to 2d quantum gravity emerged. The first one is called Liouville approach \cite{Seiberg:1990eb}, which is formulated in the continuum 2d spacetime; the second approach is based on a discretisation of 2d random surfaces and described in terms of matrix models \cite{Ginsparg:1991bi}. The classical Liouville theory was extensively studied at the end of the last century in connection with the uniformization problem for Riemann surfaces. An interesting feature of Liouville theory is that it can be quantized as a conformal field theory (CFT), implying in particular that the space of states forms a representation of the Virasoro algebra. Liouville theory seems to be a kind of universal building block for a variety of models for two dimensional gravity and non-trivial backgrounds in string theory.  On the other hand,  matrix models provide explicit non-perturbative solutions of 2d quantum gravity and/or strings in spacetimes with dimensions less than or equal to two. For a summary of the Liouville theory and the matrix models, we recommend the reader to the review \cite{Nakayama:2004vk}. 


In general relativity, there is no non-trivial gravitational dynamics in spacetime dimensions lower than four. In three dimensions, Riemann tensor is proportional to Ricci tensor and the source-free theory is trivial. In two dimensions the Einstein tensor vanishes and the Einstein-Hilbert action is a topological invariant. So the equations of motion do not exist and hence one cannot formulate a meaningful theory. This was remedied by a proposal from Teitelboim \cite{Teitelboim:1983ux} and Jackiw \cite{Jackiw:1984book}. They independently suggested that an appropriate geometrical model for two dimensional gravity is the constant curvature equation,
${}^{(2)}R - 2 \Lambda = 0$,
which is the natural analog of the vacuum Einstein equations with a cosmological constant, where ${}^{(2)}R$ denotes the two dimensional Ricci scalar of the spacetime. To quantize the theory, one would need a local action principle from which this equation can be derived. It is also expected that such an action be general covariant if it were to be useful to understand the four-dimensional gravity. They found that the only invariant action  is the non-geometric action involving a scalar field $\Phi$ as a Lagrange multiplier
$S= \int d^2x\, \Phi \sqrt{-g} \left({}^{(2)}R - 2 \Lambda\right)$,
that yields the desired equation upon varying with respect to $\Phi$. The exact solution of this model was found by Henneaux \cite{Henneaux:1985nw}.

In this paper, we examine the two-dimensional version of Ho\v{r}ava-Lifshitz (HL) theory without projectability \cite{Horava}, where we extend the  canonical quantization techniques we earlier employed in the projectable version  of the theory \cite{L3W} to the non-projectable case.  The paper is organized as follows. We give a basic set-up of the two-dimensional HL theory without projectability in Section II and discuss its classical solutions in Section III. Section IV is the heart of this paper in which we study the Hamiltonian structure and canonically quantize the theory, while in Section V we summarize our main conclusions. Before proceeding further, we note that quantization of the 2d HL theory was also studied in the framework of causal dynamical triangulations \cite{Ambjorn}. 


\section{2d non-projectable Ho\v{r}ava-Lifshitz gravity}

The general gravitational action of the HL gravity is given by,
$S_{HL}=\zeta^2 \int {dt\, dx\, N\sqrt{g} \left({\cal{L}}_{K} - {\cal{L}}_{V}\right)}$,
where   $N$ denotes  the lapse function in the Arnowitt-Deser-Misner (ADM) decomposition \cite{ADM}, and
 $g \equiv {\mbox{det}}(g_{ij})$, here $g_{ij}$ is the spatial metric defined on the leaves $t=$ Constant.   ${\cal{L}}_{K}$ is the kinetic part of the action, given by
${\cal{L}}_{K} = K_{ij}K^{ij} - \lambda K^2$,
where $\lambda$ is a dimensionless constant, and $K_{ij}$ denotes the extrinsic curvature tensor of the leaves $t=$ constant, given by 
$K_{ij}=\frac{1}{2N}\left(-\dot g_{ij}+\nabla_iN_j+\nabla_jN_i\right)$,
and   $K \equiv g^{ij}K_{ij}$. Here $\dot{g}_{ij} \equiv \partial{g}_{ij}/\partial t$, $\nabla_i$ denotes the covariant derivative of the metric $g_{ij}$, and $N^i$ the shift vector,
with $N_i \equiv g_{ij}N^j$.

On the other hand,  ${\cal{L}}_{V}$ denotes the potential part of the action, and is made of $R, \; \nabla_i$ and $a_i$, 
that is, 
${\cal{L}}_{V} = {\cal{L}}_{V}\left(R, \; \nabla_i, \; a_i\right)$,
where $a_i \equiv N_{,i}/N$ and $R$ denotes the Ricci scalar of the  leaves $t=$ Constant, which identically vanishes in one-dimension, i.e., $R = 0$. Power-counting renormalizibility 
condition requires that ${\cal{L}}_{V}$ should contain spatial operators with the highest dimensions that are not less than $2z$, where $z \ge d$ \cite{Horava,Visser}, and $d$
denotes the number of the spatial dimensions. Taking the  minimal requirement, that is, $z =d$, we find that in the current case ($d = 1$) we have 
${\cal{L}}_{V} = 2\Lambda - \beta a_i a^i$,  
where $\Lambda$ denotes the cosmological constant, and $\beta$ is another dimensionless coupling constant. 

Collecting all the above together, the gravitational action of the HL gravity in $(1+1)$-dimensional spacetimes can be cast in the form, 
\bq
\lb{2.11}
S_{HL}=\zeta^2 \int {dt\, dx\,   N\gamma \left[(1-\lambda)K^2-2{\Lambda} + \beta a_ia^i\right]},
\eq
 where $\gamma \equiv \sqrt{g_{11}},\;   \gamma' \equiv \partial\gamma/\partial x$, and
\bqn
\lb{2.8}
K&=& g^{11}K_{11} = -\frac{1}{N}\left(\frac{\dot\gamma}{\gamma}-\frac{N_1'}{\gamma^2}+\frac{N_1\gamma'}{\gamma^3}\right),
\eqn
with $N_1 \equiv g_{1i}N^i = \gamma^2N^1$. 

Regarding to the above general action (\ref{2.11}), it is interesting to note that,  in a particular gauge, the so-called $T$-gauge \cite{JacobsonA,Wang},  
in which the aether field $u_a$ can be written as
 \cite{SVW},
$u_{a} = {t_{,a}}/{\sqrt{-t_{,b}t^{,b}}}$,
where $t$ is the global time introduced  above in the HL gravity, the action of the 2d Einstein-aether theory \cite{EJ} is identical to the action (\ref{2.11}). 
It should be noted that this identification is  only in the action level,  as the two theories  have different gauge symmetries, and the 2d HL theory is  only
a gauge-fixed form of the  2d Einstein-aether one.  Contrary examples can be found in \cite{CW10,Wang}.

\section{Classical Solutions}

The line element in terms of $N, \; N^1$ and $\gamma$,  takes the form \footnote{The general classical solutions of the 2d Einstien-aether theory without the cosmological constant  $\Lambda$ were studied in detail  in \cite{EJ}.},
\bq
\lb{metric}
ds^2=-N^2(t, x)dt^2+\gamma^2(t, x)\big(dx+N^1(t, x)dt\big)^2,
\eq
with the gauge freedom,
\bq
\lb{gauge}
t' = \xi^0(t), \quad
x' = \xi^1\left(t, x\right),
\eq
where $\xi^0(t)$ and $\xi^1\left(t, x\right)$ are arbitrary functions of their indicated arguments.
Variations of the   the action Eq.(\ref{2.11})  with respect to $\gamma, \; N$ and $N_1$ yield respectively the following equations
\bqn
\lb{4.1a}
&& 2(1-\lambda)\left[\dot{K}-\frac{NK^2}{2}-\frac{KN_1'}{\gamma^2}+\frac{2KN_1\gamma'}{\gamma^3} +\left(\frac{KN_1}{\gamma^2}\right)'\right] \nb\\
&& ~~~~~~~~~~~~ ~~~~~ 
 -\frac{\beta N'^2}{N\gamma^2}-2\Lambda N =  0, \\
\lb{4.1b0}
&&  (1-\lambda) \gamma K^2+2\Lambda\gamma+2\beta\Big(\frac{N'}{N\gamma}\Big)'+\beta\frac{N'^2}{N^2\gamma} = 0,~~~~ 
\eqn
and $K' = 0$. Thus, we have $K = K(t)$.  Using the gauge freedom (\ref{gauge}), we can always set 
$N^1(t, x) = 0$,
 without loss of the generality. 
It should be noted that this gauge choice does not completely fix the gauge freedom, and the remaining one is,  
\bq
\lb{gaugeB}
t' = \xi^0(t), \quad
x' = \hat\xi^1\left(x\right).
\eq
  With  the gauge $N_1=0$, Eq.(\ref{2.8}) reduces to, 
$K(t) = - {\dot{\gamma}}/({N\gamma})$,
while Eqs.(\ref{4.1a}) and (\ref{4.1b0}) reduce to 
  \bqn
\lb{4.1aa}
 (1-\lambda)K^2-2(1-\lambda)\frac{\dot{K}}{N}+\beta y^2+2\Lambda &=& 0,\\
\lb{4.1dd}
{2}{}y'+\left(y^2 - g(t)\right)\gamma= 0,
\eqn
where $y \equiv {N'}/{(N\gamma)}$, and $g(t)\equiv -{\beta}^{-1}\left[(1-\lambda)K^2+2\Lambda\right]$.
Eq.(\ref{4.1dd}) has the general solution,  
\bqn
\lb{4.1b}
y(t, x) &=& -\sqrt{g(t)}\tanh\Delta(t, x), \nb\\
\Delta(t, x) &\equiv& -\sqrt{g(t)}\left[\frac{\int^x\gamma(t,x') dx'}{2}-c_1(t)\right], ~~~~~
\eqn
where $c_1(t)$ is an arbitrary function of $t$ only.  On the other hand, from Eqs.(\ref{4.1aa}) and (\ref{4.1dd}), we find that
\bq
\lb{4.3}
(1-\lambda)\gamma\dot{K}+\beta Ny'=0,
\eq
from which, together with Eq.(\ref{4.1b}), we find,  
$N(t, x) = N_0(t)\,  \hat{N}(t,x)$, 
where
$\hat{N}(t,x) = 2 \cosh^2\Delta(t, x)$ and $ N_0(t) =  {(\lambda-1)\dot{K}}/[{\beta g(t)}]$. 
 Using the remaining gauge freedom of Eq.(\ref{gaugeB}), we can always absorb the factor $N_0(t)$ into $t'$, so the lapse function finally takes the form,
\bqn
\lb{4.5a}
{N}(t,x) = 2\cosh^2\Delta(t, x).
\eqn
Inserting it, together with $y$ given by Eq.(\ref{4.1b}), into Eq.(\ref{4.3}) we find that 
\bq
\lb{4.6}
\dot{K}(t) -K^2(t) +\eta=0,
\eq
where
$\eta\equiv {2\Lambda}/{(\lambda-1)}$.
When $\dot{K} =0$, Eq.(\ref{4.6}) has the solution,  
$K=\pm\sqrt{\eta}$. 
Clearly, for $K$ to be real, we must assume $\eta \ge 0$. Then, from (\ref{4.5a}) we find that $g(t) = 0$ and $N(t, x) =2$. 
Redefining $t$, we can always set $N =1$. Then,   from Eq.(\ref{4.3}) we find that
$\gamma(t, x) = \gamma_0(x)e^{\mp2\sqrt{\eta}(t -t_0)}$,
where $\gamma_0(x)$ is an arbitrary function of $x$, and $t_0$ is a constant. Using the gauge residuals of Eq.(\ref{gaugeB}) we can always set $\gamma_0(x) = 1$ and $t_0 = 0$, so the
corresponding metric finally takes the form, 
\bq
ds^2=-dt^2+e^{\mp4\sqrt{\eta}t}dx^2,\; (\dot{K} =0),
\eq
which is nothing but the de Sitter spacetime.

When $\dot{K} \not=0$, Eq.(\ref{4.6}) has a solution,  
$K(t)=-\sqrt{\eta}\tanh\big[\sqrt{\eta}(t-t_0)\big]$,
 from  which we find that, $g(t)=-({2\Lambda}/{\beta)\cosh^{-2} \left[\sqrt{\eta}(t-t_0)\right]}$.
On the other hand, combining Eqs.(\ref{4.3}) and (\ref{4.5a}) we find 
\bq
\lb{gEq}
{\dot{\gamma}} + 2K(t) \cosh^2\Delta {\gamma}= 0,\; (\dot{K} \not=0),
\eq
where $\Delta(t, x)$ is given  by  Eq.(\ref{4.1b}).

\section{Hamiltonian Structure and Canonical Quantization}

Now,  let us turn to the Hamiltonian structure and canonical quantization.   For such a purpose, in this section we shall not restrict ourselves to any gauge.   Then,   from the action (\ref{2.11}) we find that the canonical momenta are given by,
\bqn
\lb{1.3}
\pi_{N}&\equiv&\frac{\partial \mathcal{L}}{\partial\dot{N}} =  0,\quad
\pi_{N_1}\equiv\frac{\partial \mathcal{L}}{\partial\dot{N_1}} =  0,\nb\\
\pi&\equiv&\frac{\partial \mathcal{L}}{\partial\dot{\gamma}}=2\zeta^2(\lambda-1) K,\nb
\eqn
with K given by Eq.(\ref{2.8}).  
After Legendre transformation, the Hamiltonian density is given by, 
\bqn
\lb{1.5}
\mathcal{H}  
&=&\frac{N\gamma \pi^2}{4\zeta^2(1-\lambda)}+2\zeta^2\Lambda\gamma N-\frac{N_1\pi'}{\gamma}\nb\\
&-&\frac{\beta\zeta^2N}{\gamma}\left(\frac{N'}{N}\right)^2+\pi_N \sigma+\pi_{N_1}\sigma_1,
\eqn
where $\sigma$ and $\sigma_1$ are the Lagrangian multipliers. Then,  the Hamiltonian takes the form,
\bq
\lb{1.6}
H=\int{dx\, \mathcal{H}(x)}.\nb
\eq
Now, the preservation of the primary constraints, $\pi_N\approx 0$ and $\pi_{N_1}\approx 0$, gives us the secondary constraints. By evaluating the poisson brackets we find 
\bqn
\lb{1.7}
\dot{\pi}_{N_1}=\Big\{ \pi_{N_1}, H\Big\}=-\mathcal{H}_1\approx 0, \nb\\
\dot{\pi}_N=\Big\{ \pi_N, H\Big\}=-\mathcal{H}_2\approx 0. \nb
\eqn
Here 
\bqn
\lb{1.8}
\mathcal{H}_1&\equiv&-\frac{\pi'}{\gamma},\\
\lb{1.81}
\mathcal{H}_2&\equiv&\frac{\pi^2\gamma}{4\zeta^2(1-\lambda)}+2\zeta^2\Lambda\gamma\nb\\
&+&\beta \zeta^2 \frac{N'^2}{N^2\gamma}+2\beta \zeta^2\left(\frac{N'}{N\gamma} \right)'.
\eqn
Rearranging the Hamiltonian in terms of the constraints, we end up with
\bqn
\lb{1.9}
\mathcal{H} &=& N_1\mathcal{H}_1+N\mathcal{H}_2+\pi_N\sigma+\pi_{N_1}\sigma_1 \nb\\
&& -2\beta\zeta^2\left(\frac{N'}{\gamma}\right)'.
\eqn
In the following  analysis, we will drop the last surface term.  
By straightforward calculations, we can obtain the structure functions of the constraints, which are given by, 
\bqn
\lb{1.10}
\Big\{\mathcal{H}_1(x),\mathcal{H}_1(x')\Big\}&=&\left(\frac{\mathcal{H}_1(x')}{\gamma^2(x')}+\frac{\mathcal{H}_1(x)}{\gamma^2(x)}\right)\partial_{x'}\delta(x-x'), \nb\\
\lb{1.10a}
\Big\{\mathcal{H}_1(x),\mathcal{H}_2(x')\Big\}&=&-\frac{\pi(x)\mathcal{H}_1(x)}{\zeta^2(1-\lambda)}\delta(x-x')\nb\\
&+&\frac{\mathcal{H}_2(x)}{\gamma^2(x)}\partial_{x}\delta(x-x')\nb\\
&+& \frac{2\beta\zeta^2N'}{\gamma^3N}\partial_{xx}\delta(x-x')\nb\\
&-&2\beta\zeta^2\left(\frac{N'\gamma'}{N\gamma^4}+\frac{N'^2}{N^2\gamma^3}\right)\partial_{x}\delta(x-x')\nb\\
&-&\frac{\beta\zeta^2}{\gamma}\left(\frac{N'^2}{\gamma^2N^2}\right)'\delta(x-x').
\eqn
Clearly, $\mathcal{H}_1$ and $\mathcal{H}_2$ don't commute with each other on the constraint surface due to the last three terms in the right-hand side of
Eq.(\ref{1.10a}) (all the functions at the right-hand side of this commutator are functions of x).
In addition,  we also have    
\bqn
\lb{1.10.1}
\Big\{\mathcal{H}_2(x),\mathcal{H}_2(x')\Big\}&=&-\frac{2\beta \pi(x) N'(x)}{(1-\lambda)N(x)\gamma(x)}\partial_{x}\delta(x-x') \nb\\
&-&\frac{\beta}{1-\lambda}\left(\frac{N'\pi}{N\gamma}\right)'\delta(x-x'), \nb\\
\Big\{\pi_N(x),\mathcal{H}_2(x')\Big\}&=&-\frac{2\beta\zeta^2}{N(x')\gamma(x')}\partial_{x'x'}\delta(x-x')\nb\\
&-&2\beta\zeta^2\partial_{x'}\left(\frac{1}{N(x')\gamma(x')}\right)\partial_{x'}\delta(x-x')\nb\\
&+&\frac{2\beta\zeta^2}{N}\left(\frac{N'}{N\gamma}\right)'\delta(x-x').\nb
\eqn
So, $\pi_N$ and $\mathcal{H}_2$ don't commute either. In this case, we need to define a new constraint via the relation, 
\bq
\lb{H1}
\tilde{\mathcal{H}}_1=\mathcal{H}_1+\frac{N'}{\gamma^2}\pi_N.
\eq
As it turns out that $\tilde{\mathcal{H}}_1$ commutes with both $\mathcal{H}_2$ and $\pi_N$ on the constraint surface, and their structure functions are given by, 
\bqn
\lb{1.10.2}
\Big\{\tilde{\mathcal{H}}_1(x),\mathcal{H}_2(x')\Big\}&=&-\frac{\pi(x)\tilde{\mathcal{H}}_1(x)}{\zeta^2(1-\lambda)}\delta(x-x') \nb\\
&+&\frac{\mathcal{H}_2(x)}{\gamma^2(x)}\partial_{x}\delta(x-x'),\nb\\
\Big\{\tilde{\mathcal{H}}_1(x),\pi_N(x')\Big\}&=&\frac{\pi_N(x)}{\gamma^2(x)}\partial_x\delta(x-x'), \nb\\
\Big\{\tilde{\mathcal{H}}_1(x),\tilde{\mathcal{H}}_1(x')\Big\}&=&\left(\frac{\tilde{\mathcal{H}}_1(x')}{\gamma^2(x')}+\frac{\tilde{\mathcal{H}}_1(x)}{\gamma^2(x)}\right)\partial_{x'}\delta(x-x').\nb 
\eqn
Correspondingly, the Hamiltonian now takes the form \footnote{Hamiltonian structure of 4-dimensional HL theory without the projectability condition was studied in \cite{DJ}, and
a similar structure was obtained (See also \cite{Kluson}). We thank T. Jacobson for pointing this out to us.}, 
\bqn
\lb{1.11}
\tilde{\mathcal{H}} =N_1\tilde{\mathcal{H}}_1 + \sigma_1\pi_{N_1} +N\mathcal{H}_2+ \sigma\pi_N.
\eqn
Then, one can show that  $\pi_{N_1} \approx 0$ and $\tilde{\mathcal{H}_1} \approx 0$ are the first-class constraints, while  $\pi_N \approx 0$ and $\mathcal{H}_2 \approx 0 $ are the second-class constraints. These constraints are preserved under time evolution. So,
the physical degrees ($\mathcal{N}$) of freedom of the theory per spacetime point    is given by 
\bqn
\lb{ND}
\mathcal{N}&=&\frac{1}{2}\big({\mbox{dim}}\mathcal{P}-2\mathcal{N}_1-\mathcal{N}_2\big),\nb\\
&=&\frac{1}{2}\big(6-2*2-2\big)=0.\nb
\eqn
Here ${\mbox{dim}}\mathcal{P}$ means the dimension of the phase space, and $\mathcal{N}_1$($\mathcal{N}_2$) denotes the number of first-class (second-class) constraints. It is interesting to note that $\mathcal{N}$
is not equal to $-1$, as in the usual 2d relativistic case \cite{Henneaux:1985nw}, due to the new gauge symmetry (\ref{gauge}) of the theory.
It is also interesting to note that in the projectable case the  physical degrees of freedom is also zero \cite{L3W}.

Now we proceed to the canonical quantization of the system by following  Dirac \cite{Dirac}. First,  for the two second-class constraints $\pi_N\approx 0$ and $\mathcal{H}_2\approx 0$, we can make them strongly equal to zero,
\bq
\lb{strong}
(i) \;\; \pi_N = 0, \quad (ii)\;\;  \mathcal{H}_2 = 0,
\eq
 by simply adopting Dirac's bracket,  instead of the Poisson one.  The first condition is actually empty, while from the second condition  we can express  $N$ as a functional of $\gamma$ and $\pi$ by solving the equation  $\mathcal{H}_2=0$, where $\mathcal{H}_2$ is given by Eq.(\ref{1.81}). The general  solution is given by
\bq
\lb{Nf}
N(t,x) = N_0(t)\exp\left\{{\int^{x}{y(t, x')\gamma(t, x') dx'}}\right\},
\eq
where $N_0(t)$ is an integration function of $t$ only, and  $y(t, x)$ is given by Eq.(\ref{4.1b}).  As a result,  we can drop $N$ and $\pi_N$ by going to the ``reduced" phase space spanned by   $\left(N_1, \pi_{N_1}; \gamma, \pi\right)$. However, the phase space can be further reduced
 by noting that the first-class constraint $\pi_{N_1}\approx0$ simply yields,
 \bq
 \lb{WQa}
 - i \hbar \frac{\delta\psi}{\delta N_1}  = 0,\nb
 \eq
 that is, the wavefunction $\psi$ will not depend on $N_1$ and $\pi_{N_1}$. Then, the reduced phase actually becomes two-dimensional, spanned by $\gamma$ and $\pi$. 
 
 On the other hand, with the first condition (\ref{strong}), the first-class 
 constraint $\tilde{\mathcal{H}_1} \approx 0$ reduces to ${\mathcal{H}_1} \approx 0$, as one can see from Eq.(\ref{H1}). This in turn implies, 
\bq
\lb{1.14}
\pi - \alpha(t) \approx 0, \nb
\eq
where $\alpha(t)  \equiv 2\zeta^2(\lambda-1) K(t)$. Then, the corresponding Wheeler-DeWitt equation takes the form,  
\bq
\lb{eqwf}
\left(-i\hbar \frac{\delta}{\delta\gamma} - \alpha(t)\right)\psi\left(\gamma; t\right) =  0,
\eq
which has the general (plane-wave)  solution,  
\bq
\lb{wf}
\psi\left(\gamma, t\right)=\psi_0 e^{\frac{i\alpha}{\hbar}L}.
\eq
Here,  $L \equiv L(t)$ is the gauge-invariant length, defined as  \cite{L3W}, 
\bq
\lb{1.15}
L(t) \equiv \int_{-L_{\infty}}^{L_{\infty}}{\gamma(t, x) dx},
\eq
where $x = \pm L_{\infty}$ represent the boundaries of the one-dimensional spatial space. The   integration ``constant" $\psi_0$  in general is a function of $t$. But, the normalization condition,
$ \int_{-L_{\infty}}^{L_{\infty}}{\left|\psi\right|^2 dx}$ requires $\psi_{0} = e^{i\beta(t)}/(2L_{\infty})$, where $\beta(t)$ is real and otherwise  arbitrary function of $t$ only. However, without loss of the generality,
we can always set $\beta(t) = 0$.

\section{Conclusions}

In this paper, we have studied the quantization of 2d  Ho\v{r}ava theory of gravity without the projectability condition, that is, the lapse function $N$ in general is a function of both time  and space, $N = N(t, x)$.  The classical solutions have been studied in some detail  and shown that the extrinsic curvature of the leaves $t = $ Constant  is always independent of the spatial coordinate.   In the case of a constant extrinsic curvature the corresponding spacetime is de Sitter, while in the general case,  the dynamical variable $\gamma(t, x)$ satisfies  a master equation given by Eq.(\ref{gEq}). Once $\gamma$ is known, the rest of metric coefficients can be found algebraically. 

Our investigation of the Hamiltonian structure of the theory shows that the system consists of two first-class and two second-class constraints. As a result, the number of total degrees of freedom is zero. Following Dirac \cite{Dirac}, we have first turned the two second-class constraints into strong ones, by requiring that they be strongly equal to zero, from which we can express the lapse function $N$ as a functional of the canonical  variable $\gamma$ and its momentum conjugate  $\pi$, so the phase space is reduced from six-  to four-dimensions, spanned by $\left(N_1, \pi_{N_1}; \gamma, \pi\right)$. But, one of the two first-class constraints further tells us that the actual dimension of the phase space is two, since the wavefunction of the system is independent of the shift vector $N_1$ and its momentum conjugate $\pi_{N_{1}}$. As a result, the corresponding Wheeler-DeWitt equation simply takes the form of Eq.(\ref{eqwf}) and has a plane-wave solution (\ref{wf}), in terms of the gauge-invariant length $L(t)$ defined by Eq.(\ref{1.15}). Therefore, similar to the projectable case \cite{L3W}, this system  is also  quantum mechanical in nature. This is understandable, as  this system too has zero-degree of freedom. However, what is a bit surprising is that the corresponding Wheeler-DeWitt equation simply yields the plane-wave solution.

 In addition, the classical spacetimes do not play important role in  the process of quantization. In particular,  it does not matter whether the classical background is de Sitter or not, the wavefunction is  always a plane-wave solution. The only effects of the classical backgrounds are encoded in   the phase of the plane-wave, in terms of  the   extrinsic curvature $K(t)$ of the leaves $t = $Constant, where $t$ is the time coordinate, with which the spacetime is foliated  globally.

\section*{Acknowledgements}

 We would like to thank Ted Jacobson for valuable suggestions/comments. Part of the work was done when A.W. and B.-F.L. were visiting Zhejiang University of Technology (ZUT), China, and part of was done when A.W.
 was visiting  the State University of Rio de Janeiro (UERJ), Brazil. They would like to thank ZUT and UERJ for their hospitality.  This work is supported in part by Ci\^encia Sem Fronteiras, 
 No. 004/2013 - DRI/CAPES, Brazil (A.W., V.H.S.); Chinese NSFC No. 11375153 (A.W.) and No. 11173021 (A.W.). B.-F.L. is supported by Baylor University through the physics graduate programs.

\end{document}